\documentclass[a4,14pt]{article}
\usepackage[margin=1in]{geometry}
\usepackage{subcaption}
\usepackage{amsmath}
\usepackage{hyperref}
\usepackage{amssymb}
\usepackage{braket}
\usepackage{upgreek}
\usepackage{graphicx}
\usepackage{float}
\usepackage{cite}
\begin{document}
\pagenumbering{arabic}
\title{Chameleon-neutrino conformal coupling and MSW-mediated solar neutrino deficit}
\author{H. Yazdani Ahmadabadi\footnote{hossein.yazdani@ut.ac.ir} and H. Mohseni Sadjadi\footnote{mohsenisad@ut.ac.ir}
\\ {\small Department of Physics, University of Tehran,}
\\ {\small P. O. B. 14395-547, Tehran 14399-55961, Iran}}
\maketitle
\begin{abstract}
A modified version of the MSW effect is studied within the framework of screening models through conformal couplings of a scalar field within matter.
Such a coupling leads to a new, unknown interaction between the scalar field and the neutrino.
We mainly look for the discrepancy in the total solar electron-neutrinos flux through this interaction.
Since the scalar field behavior depends on the local density of matter, we observe an indirect effect of matter on flavor changes, and subsequently, on neutrino flux through neutrino decay.
The paper ends by describing various probabilities and comparing the results with observational data.
\end{abstract}

\section{Introduction}\label{sec1}
The positive accelerated expansion of the Universe caused by a dynamical scalar field is among the biggest mysteries of cosmology.
It seems that there is no obvious evidence for such a component in the local observations and experiments.
To pass local gravitational tests, a ``screening mechanism'' is needed to suppress the scalar field on small scales and produce observable signals on cosmological scales \cite{KhouryPRL,Khoury-Weltman,Brax1,Brax2,Waterhouse,Tsujikawa,Burrage,Elder,Hinterbichler-SYMM1,Hinterbichler-SYMM2,MohseniSadjadi-SYMM,Honardoost-SYMM}.
Among the various screening proposals, a quintessential field interacts with matter components through a conformal coupling \cite{Bean, MohseniSadjadi-Honardoost, Honardoost-Sadjadi}.
This type of coupling may give rise to the Chameleon model \cite{KhouryPRL,Khoury-Weltman,Brax1,Brax2,Waterhouse,Tsujikawa,Burrage,Elder}, in which the scalar field's mass is an increasing function of the ambient matter density that allows the field to screen itself in dense regions.

Motivated by the approximate equivalence of the neutrino mass and the dark energy scale (of order $\sim$ meV), a scalar-neutrino non-standard interaction (NSI) study \cite{MSWWolf, Miranda, FengGe, BhupalDev} has recently been used to constrain the exotic field's properties \cite{Amendola, Brookfield, CarrilloGonzalez, Mandal, Salazar-Arias,Khalifeh}.
So, in addition to the traditional probes for the screening models, such as the E$\ddot{\text{o}}$t-Wash \cite{Will-GR.EXP,Nobili-WEP} and atom interferometry \cite{Interfero1-Hamilton,Interfero2-Elder,Interfero3-Sabulsky,Interfero4-Lemmel}, looking at neutrinos could open a new window to the nature of dark energy.
Most of the observational neutrino flavor conversion data from experiments such as Super-Kamiokande (SK) \cite{Super-Kamiokande:2002weg}, Sudbury Neutrino Observatory (SNO) \cite{SNO:2002tuh}, and KamLAND \cite{KamLAND} can be accommodated and analyzed, showing a deficit in the total flux of detected neutrinos \cite{Deficit-McDonald,Deficit-Kajita1,Deficit-Kajita2,Deficit-Bellerive}.
Along their path towards detectors, neutrinos propagate through the vacuum or matter.
For neutrinos traveling in the vacuum, the Schr$\ddot{\text{o}}$dinger-like equation describing neutrino oscillations can be solved precisely, resulting in the phase of oscillations \cite{Curved.Cardall,Curved.Fornengo,Curved2.Visinelli,Chakraborty1,Chakraborty2,Curved4,Curved5,Curved6,Curved7.Sadjadi,Curved8.Yaz}.
However, when neutrinos propagate in the matter, neutrino flavor change is affected by their forward elastic scattering from solar electrons (i.e., caused by the standard weak interactions).
An effective matter potential describes its effect due to the Mikheyev-Smirnov-Wolfenstein (MSW) effect \cite{MSWWolf, MSWMS},
which is an adiabatic flavor conversion in a medium with slowly varying mass density and can dramatically impact the neutrino oscillation phenomenon \cite{Zhang}.

In neutrino mass varying models \cite{Fardon, Gu, Barger, Cirelli, Wetterich, MohseniSadjadi-Anari1, MohseniSadjadi-Anari2}, the neutrinos and a scalar field are linked by the neutrino mass, which has roots in the dark energy model.
The dependence of the neutrino mass on the environment due to the interactions with dark sectors demonstrates that the NSI between the scalar field and matter components modifies the MSW effect, and subsequently, generates damping signatures \cite{Curved7.Sadjadi,Kaplan}, which triggers the idea that a second-order effect might contribute to the experimentally confirmed deficit of neutrino flux.

In the present work, the scalar-matter (including neutrinos) conformal coupling manifests itself through an exponential damping factor in the conversion probabilities, which explains the MSW-mediated deficit of the neutrino flux. 
Along with neutrino oscillation, ``neutrino decay'' as a subdominant solution to explain the depletion of the solar neutrinos has been considered \cite{Bahcall}. This study will assume the heavier mass eigenstates to be unstable and then, obtains bounds on the coupling parameter.

The outline of the paper is as follows:
In section \ref{sec2}, we develop the idea of neutrino flavor change affected by neutrino-scalar field conformal coupling starting with the neutrino action, including the interacting part inside matter and the MSW effect.
With the contribution from the NSI added to the usual MSW Hamiltonian, we explicitly calculate the effective scalar-dependent mass, the wavefunction, and the weak Fermi parameter.
We then obtain effective mass and mixing parameters by diagonalizing the effective neutrino Hamiltonian.
We will discuss the possibility of neutrino decay and deficit in the total probability in section \ref{sec3}.
Furthermore, we obtain various exact damped transition probabilities in this section.
Results and their discussions are presented in section \ref{sec4}.
We will compare our results to the Borexino \cite{Agostini} and SNO+SK \cite{Zyla-PDG} data for the large mixing angle (LMA) MSW survival probability.
Concluding remarks have been brought in section \ref{sec5}.

Throughout this paper we will use units $\hbar =c= 1$, and metric signature $(-,+,+,+)$.

\section{Modified MSW effect and damping signatures}\label{sec2}
The standard neutrino interaction is associated with $\nu_e$-$e^-$ interactions through the charged weak currents.
Improved experimental precision may reveal the effects of physics beyond standard model, including NSIs of neutrino.
These non-standard couplings to a scalar field can modify the effective neutrino mixing parameters and consequently transition probabilities inside matter.
We start by considering an action in which a scalar field $\phi$ couples to different matter species $\Psi_i$ by definition \cite{Faraoni:1998qx,Carneiro:2004rt}
\begin{eqnarray}\label{eqn1}
\tilde{\text{g}}^{(i)}_{\mu\nu} = A_i^2(\phi) \text{g}_{\mu\nu},
\end{eqnarray}
that is,
\begin{eqnarray}\label{eqn2}
S_{\phi}= \int d^4x \sqrt{-\text{g}}\bigg[\frac{M_{p}^2}{2} \mathcal{R} - \frac{1}{2} \text{g}^{\mu\nu} \partial_\mu \phi \partial_\nu \phi - \mathcal{V}(\phi) \bigg] + \int d^4x  \mathcal{L}_m \big(\Psi_i , \tilde{\text{g}}^{(i)}_{\mu\nu}\big),
\end{eqnarray}
where $\text{g}$ is the determinant of the metric $\text{g}_{\mu\nu}$, $M_{p}$ is the reduced Planck mass, $\mathcal{R}$ is the Ricci scalar, $\mathcal{V}(\phi)$ is the scalar field potential, and $\mathcal{L}_m$ is the Lagrangian density of matter components.
In this work, we assume that the matter is universally coupled to the metric $\tilde{\text{g}}_{\mu\nu}$.
On the other side, we can generalize the model to the various arbitrary functions $A_{i}(\phi)$ for each matter component, which leads us to the violation of the weak equivalence principle (WEP) in cosmological scales \cite{Khoury-Weltman, Waterhouse, Burrage}.
We write the coupling function $A(\phi)$ in an exponential form, i.e., we have
\begin{eqnarray}\label{eqn3}
A(\phi) = \exp\left[\frac{\beta \phi}{M_p}\right],
\end{eqnarray}
where $\beta$ denotes the matter-scalar coupling strength.
Theories with $\beta \sim 1$ have gravitational strength couplings to matter \cite{Burrage}, whereas the $\beta \to 0$ reproduces the no-Chameleon case.

In general, NSIs can be considered to be effective additional contributions to the standard vacuum action that describes the neutrino evolution.
Including the $\nu_e\text{-}e^-$ interacting term, we have the following ``fermionic'' action in the original frame (here, the tilde frame) \cite{Cottingham}:
\begin{eqnarray}\label{eqn4}
\begin{split}
&\tilde{S}_{\nu}^{\text{m}}  = \int d^4x \sqrt{-\tilde{\text{g}}} \bigg[- \sum_{\alpha,\beta} \upnu^{\dagger}_{\alpha L} m_{\alpha \beta} \upnu_{\beta R} - \sum_{\alpha,\beta} \upnu^{\dagger}_{\alpha R} m_{\beta \alpha}^{*} \upnu_{\beta L} \\& ~~~~ + i \sum_\alpha
\left(\upnu_{\alpha L}^{\dagger} \tilde{\sigma}^\mu_L \tilde{D}_\mu \upnu_{\alpha L}  + \upnu_{\alpha R}^{\dagger} \tilde{\sigma}^\mu_R \tilde{D}_\mu \upnu_{\alpha R} \right) - 2\sqrt{2}G_F \left(e_L^{\dagger} \tilde{\sigma}^\mu_L \upnu_{eL}\right)\left(\upnu^{\dagger}_{eL} \tilde{\sigma}_{\mu L} e_L\right)\bigg],
\end{split}
\end{eqnarray}
where $G_F$ is the weak Fermi constant, $\tilde{D}_\mu$ is the covariant derivative in the original frame, and $\tilde{\sigma}^\mu_R \equiv \left(\tilde{\sigma}^0,\tilde{\sigma}^1,\tilde{\sigma}^2,\tilde{\sigma}^3\right)$ and $\tilde{\sigma}_L^\mu \equiv \left(\tilde{\sigma}^0,-\tilde{\sigma}^1,-\tilde{\sigma}^2,-\tilde{\sigma}^3\right)$ are respectively the Pauli matrices for the right- and left-handed spinors in the original frame.
It is to be noted that the conformal transformation from the original frame (tilde frame) to a new frame is sometimes applied as a purely mathematical tool to handle problems by reducing the model to a familiar and computationally more convenient scenario.
After calculations, we bring back all the quantities to the original frame, wherever needed.

Using the Fierz transformation \cite{FierzTransformation}, a neutral current interaction can give the interacting part of the above action. Hence, we have
\begin{eqnarray}\label{eqn5}
\tilde{\mathcal{L}}_{\text{int.}} = -2\sqrt{2}G_F \eta_{\hat{a} \hat{b}} \left(e_L^{\dagger} \sigma^{\hat{a}}_L e_{L}\right)\left(\upnu^{\dagger}_{eL} \sigma^{\hat{b}}_L \upnu_{eL}\right).
\end{eqnarray}
Note that we have used the fact that $\tilde{\sigma}_L^{\mu} = \epsilon^\mu_{\hat{a}} \sigma_L^{\hat{a}}$, where $\epsilon_\mu^{\hat{a}}$ is the tetrad field.
For the solar medium electrons at rest (i.e., non-relativistic), we have
\begin{eqnarray}\label{eqn6}
e^{\dagger}_L \sigma^{\hat{0}}_L e_L = e^{\dagger}_L e_L = \frac{1}{2} N_e(z),
\end{eqnarray}
where $N_{e}(z)$ is the number density of electrons inside matter, and the factor $1/2$ implies that we only consider the left-handed electrons.
It is easy to check out that the expectation value of $e^{\dagger}_L \sigma_L^{\hat{i}} e_L$ is zero.

We predict that the conformal coupling of the neutrino-scalar field can impact the neutrino characteristics and, hence, modifies the flavor conversion probabilities. 
It will be useful to remember the transformation properties of tetrads, gamma matrices, and Christoffel connections \cite{Curved7.Sadjadi}.
By using the properties of tetrads
\begin{eqnarray}\label{eqn7}
	\tilde{\epsilon}_{\mu}^a = A(\phi)\epsilon_{\mu}^a,
\end{eqnarray}
gamma matrices
\begin{eqnarray}\label{eqn8}
	\tilde{\gamma}^\mu = A^{-1}(\phi) \gamma^\mu,
\end{eqnarray}
and Christoffel connections
\begin{eqnarray}\label{eqn9}
	\tilde{\Gamma}_{\mu\nu}^{\lambda} = \Gamma_{\mu\nu}^{\lambda} - A^{-1}(\phi) A^{,\lambda}(\phi)g_{\mu\nu} + A^{-1}(\phi) A_{,\mu} (\phi) \delta_{\nu}^{\lambda} + A^{-1}(\phi) A_{,\nu}(\phi) \delta_{\mu}^{\lambda}
\end{eqnarray}
under the conformal transformation (\ref{eqn1}), it can be easily demonstrated that
\begin{eqnarray}\label{eqn10}
	\tilde{\gamma}^\mu \tilde{D}_\mu = A^{-1}(\phi) \gamma^\mu D_\mu + \dfrac{3}{2} A^{-2}(\phi) A_{,\mu}(\phi) \gamma^\mu.
\end{eqnarray}
The non-standard coupling term embedded in the covariant derivative is added to the fermionic part action in Eq.(\ref{eqn4}). The action (\ref{eqn4}) may remain invariant in the new frame, i.e.,
\begin{equation}\label{eqn11}
\begin{split}
& \int d^4x \sqrt{-\tilde{\text{g}}}~\left[\bar{\upnu}(x)\left(i\tilde{\gamma}^\mu \tilde{D}_\mu - m \right)\upnu(x) - \sqrt{2} N_e G_F \bar{\upnu}(x) \upnu(x)\right] \\& = \int d^4x \sqrt{-\text{g}}\left[\bar{\upnu^\prime}(x) \left(i\gamma^\mu D_\mu - m^\prime\right)\upnu^\prime(x) - \sqrt{2} N_e G^\prime_F \bar{\upnu}^\prime(x) \upnu^\prime(x)\right],
\end{split}
\end{equation}
provided that the scalar-dependent effective neutrino mass
\begin{eqnarray}\label{eqn12}
m^\prime (x) = A(\phi) m,
\end{eqnarray}
the wavefunction
\begin{equation}\label{eqn13}
\upnu^\prime(x) = A^{\frac{3}{2}}(\phi) \upnu(x),
\end{equation}
and the weak Fermi parameter
\begin{eqnarray}\label{eqn14}
G_F^\prime (x) \equiv A(\phi) G_F,
\end{eqnarray}
with NSIs during propagation processes in matter are derived, where hereinafter, quantities with prime denote rescaled ones in the new frame.
We note that $\tilde{\text{g}} = A^{2d}(\phi) \text{g}$ is the determinant of the tilde spacetime metric and $d$ is the number of dimensions. For future considerations, we fix $\text{g}_{\mu\nu}$ as the Minkowski metric $\eta_{\mu\nu}$.
Also, note that the new form of the adjoint spinor $\bar{\upnu}^\prime$ is always the same as the spinor $\upnu^\prime$.
As mentioned before, we would like to work in new flat frame to handle the Dirac equation in more convenient manner with rescaled quantities above. Then, we go back to the original frame.

After applying the conformal transformation, the neutrino action in the conformally flat spacetime is therefore written as
\begin{eqnarray}\label{eqn15}
\begin{split}
&S^{\text{m}}_{\nu} = \int d^4x \sqrt{-\text{g}} \bigg[- \sum_{\alpha,\beta} \upnu^{\prime\dagger}_{\alpha L} m^\prime_{\alpha \beta} \upnu^\prime_{\beta R} - \sum_{\alpha,\beta} \upnu^{\prime\dagger}_{\alpha R} m^{\prime *}_{\beta \alpha} \upnu^\prime_{\beta L} \\& ~~~~ + i\sum_\alpha
\left(\upnu_{\alpha L}^{\prime\dagger} \sigma^\mu_L \partial_\mu \upnu^\prime_{\alpha L}  + \upnu_{\alpha R}^{\prime\dagger} \sigma^\mu_R \partial_\mu \upnu^\prime_{\alpha R} \right) - V^\prime(z) \upnu_{eL}^{\prime\dagger} \upnu^\prime_{eL}\bigg],
\end{split}
\end{eqnarray}
where $V^\prime(z) = \sqrt{2} N_e(z) G_F^\prime(z)$ is the matter potential affected by both SI and NSI of neutrinos.
Consider, for instance, the coupling function $A(\phi) = e^{\omega(\phi)}$ for the Chameleon screening model.
We can expand the exponential factor in (\ref{eqn14}) to the first order by assuming $\omega(\phi) \ll 1$, which gives
\begin{eqnarray}\label{eqn16}
\mathcal{L}^\prime_{\text{int.}} = -\sqrt{2} G_F N_e(z) \left[\upnu_{eL}^{\prime\dagger} \upnu^\prime_{eL}\right] - \sqrt{2} G_F N_e(z) \left[\omega(\phi) \upnu_{eL}^{\prime\dagger} \upnu^\prime_{eL}\right].
\end{eqnarray}
The first term is due to the SI, and the second one with $\omega(\phi) \propto \phi$ denotes a Yukawa NSI between scalar field and neutrinos.

Following the procedure in Ref.\cite{Curved7.Sadjadi} and by taking variations of the action (\ref{eqn15}) with respect to the right- and left-handed spinors
\begin{eqnarray}\label{eqn17}
	\upnu^\prime_{\alpha L}(z,t) = e^{-iE_\nu(t-t_0)}e^{+iE_\nu(z-z_0)} f^\prime_{\alpha}(z)
	\begin{pmatrix}
		0\\
		1\\
	\end{pmatrix}
\end{eqnarray}
and
\begin{eqnarray}\label{eqn18}
	\upnu^\prime_{\alpha R}(z,t) = e^{-iE_\nu(t-t_0)}e^{+iE_\nu(z-z_0)} g^\prime_{\alpha}(z)
	\begin{pmatrix}
		0\\
		1\\
	\end{pmatrix},
\end{eqnarray}
we obtain the corresponding Schr$\ddot{\text{o}}$dinger-like equation of motion in the flavor eigenbasis for neutrinos traveling inside matter:
\begin{eqnarray}\label{eqn19}
i \frac{df^\prime_{\beta}(z)}{dz} = U_{\alpha i} \left(\frac{m_{i}^{\prime 2}}{2E_\nu}\right) U_{\beta i}^* f^\prime_{\alpha}(z) + V^\prime(z) \delta_{\beta e} f^\prime_{e}(z),
\end{eqnarray}
where $f^\prime(z)$ and $g^\prime(z)$ are wavefunctions' amplitudes.
Components $U_{\alpha i}$'s are various elements of a unitary matrix called PMNS (Pontecorvo–Maki–Nakagawa–Sakata) matrix:
\begin{eqnarray}\label{eqn20}
U =
\begin{pmatrix}
c_{12} c_{13} & s_{12} c_{13} & s_{13}e^{-i\delta_{\text{CP}}}\\
-s_{12} c_{23} - c_{12} s_{23} s_{13} e^{i\delta} & c_{12} c_{23} - s_{12} s_{23} s_{13} e^{i\delta} & s_{23} c_{13}\\
s_{12} s_{23} - c_{12} c_{23} s_{13} e^{i\delta} & -c_{12} s_{23} -s_{12} c_{23} s_{13} e^{i\delta} & c_{23} c_{13} \\
\end{pmatrix}
,
\end{eqnarray}
where $c_{ij} \equiv \cos \theta_{ij}$ and $s_{ij} \equiv \sin \theta_{ij}$, and $\delta_{\text{CP}}$ is the Dirac CP-violating phase.
We can also write the Eq.(\ref{eqn19}) in the mass eigenbasis from the relation $f^\prime_i(z) = U_{\alpha i} f^\prime_{\alpha}(z)$. So, we have
\begin{eqnarray}\label{eqn21}
i \frac{df^\prime_i(z)}{dz} = \frac{m_i^{\prime 2}}{2E_\nu} f^\prime_i(z) + V^\prime(z) U_{ei} U_{ej}^* f^\prime_{j}(z).
\end{eqnarray}
Correspondingly, applying the transformation rule (\ref{eqn13}) for the neutrino wavefunction (\ref{eqn17}) results in $f^\prime_{\alpha} (z) = A^{3/2}(\phi) f_{\alpha}(z)$ in the flavor eigenbasis, or equivalently $f^\prime_{i} (z) = A^{3/2}(\phi) f_{i}(z)$ in the mass eigenbasis.
Substituting this result in the equation (\ref{eqn21}) brings us back to the original frame:
\begin{eqnarray}\label{eqn22}
i \frac{df_i(z)}{dz} = \frac{m_i^{\prime 2}}{2E_\nu} f_i(z) + V^\prime(z) U_{ei} U_{ej}^* f_{j}(z) - \frac{3}{2}i \left(\frac{d\ln A_i}{dz}\right)f_i(z) .
\end{eqnarray}
It is simply realized that the effective neutrino Hamiltonian (right-hand-side of Eq.(\ref{eqn22})) contains both the real and imaginary parts, which are respectively responsible for the modified flavor conversion inside matter (the first two terms) and damping behavior caused by NSI (the third term).
So, we have
\begin{eqnarray}\label{eqn23}
f_i(z) = e^{-i\varphi_{i}(z)} \sqrt{\mathcal{D}_{i}(z)} f_i (z_0),
\end{eqnarray}
where the quantity
\begin{eqnarray}\label{eqn24}
\mathcal{D}_{i}(z) \equiv \big[A_i[\phi(z)] A_{i}^{-1} [\phi(z_0)]\big]^{-3}
\end{eqnarray}
is defined, as a damping or an enhancing factor depending on the treatment of the scalar field $\phi$: ascending $\phi$ implies the damping factor, and descending field corresponds to the enhancing factor (for notational convenience, we will denote this by $\mathcal{D}$-factor).
We have also defined the modified scalar-dependent phase of the neutrino wavefunction
\begin{eqnarray}\label{eqn25}
\varphi_i(z) \equiv \int_{z_0}^z \mathcal{H}^{\text{Real}}_{\text{{eff.}}} d\text{z} = \frac{1}{2E_\nu}\int_{z_0}^z \left[m_i^{\prime 2} (\text{z}) + 2E_\nu V^\prime(\text{z})\right]d\text{z}.
\end{eqnarray}
Since the goal is to derive exact relations for different damped transition probabilities (in the next section), the phase part of Eq.(\ref{eqn23}) will be averaged out for very fast oscillations inside matter, which causes the oscillation phase to be disappeared in the final expressions.
Whereas, the real part of the effective Hamiltonian $\mathcal{H}^{\text{Real}}_{\text{eff.}}$ will be used to derive the effective mass and mixing parameters inside matter.
Therefore, according to Eq.(\ref{eqn17}), the normalized spacetime part of the neutrino wavefunction in the mass eigenbasis can be given by
\begin{eqnarray}\label{eqn26}
	\begin{split}
&\upnu_i(z,t) = e^{-iE_\nu\left[(t-t_0)-(z-z_0)\right]} e^{-i\varphi_{i}(z)} \sqrt{\mathcal{D}_{i}(z)} \upnu_i (z_0,t_0)
\\& ~~~~~~~~~\equiv \mathcal{F}_i(z,t)\upnu_i (z_0,t_0).
	\end{split}
\end{eqnarray}
We note that the initial value is satisfied for this relation, i.e., $\mathcal{F}_i(z_0,t_0) = 1$.

Without loss of generality, we focus on the real part of the effective neutrino Hamiltonian.
Working at the $\theta_{\text{polar}}=0$ plane, i.e., $z=r$, and with $U$ as in Eq.(\ref{eqn20}), the equation (\ref{eqn22}) in the vacuum mass eigenbasis can be given by
\begin{eqnarray}\label{eqn27}
i \frac{d}{dr}
\begin{pmatrix}
f_1\\
f_2\\
f_3
\end{pmatrix}
= \frac{1}{2E_\nu}
\begin{pmatrix}
2E_\nu V^\prime(r) c_{12}^2 c_{13}^2 & 2E_\nu V^\prime(r) c_{12} s_{12} c_{13}^2  & 2E_\nu V^\prime(r) c_{12} c_{13} s_{13}\\
2E_\nu V^\prime(r) c_{12} s_{12} c_{13}^2 & 2E_\nu V^\prime(r) s_{12}^2 c_{13}^2 + \Delta m_{21}^{\prime 2} &  2E_\nu V^\prime(r) s_{12} c_{13} s_{13}\\
2E_\nu V^\prime(r) c_{12} c_{13} s_{13} &  2E_\nu V^\prime(r) s_{12} c_{13} s_{13} & 2E_\nu V^\prime(r) s^2_{13} + \Delta m^{\prime 2}_{31}
\end{pmatrix}
\begin{pmatrix}
f_1\\
f_2\\
f_3
\end{pmatrix}
.
\end{eqnarray}
To determine the effective mass and mixing parameters induced by the neutrino-scalar NSI inside matter, we should obtain the eigenvalues and eigenvectors of the above Hamiltonian matrix.
A typical approximation that successfully describes many experiments with good accuracy is the two-flavor case.
In addition, the evolution of solar neutrinos within the Sun satisfies the condition $|\Delta m^{\prime 2}_{31}| \gg 2E_\nu V^\prime(r)$, which means neutrino flavor transition can be governed by an effective two-neutrino approach, where the evolution of a third eigenstate decouples from the other two \cite{Agarwalla,Martinez-Mirave:2021cvh}.
In the next section, we will concentrate on the various probabilities on the Earth, where the detectors have been placed.

\section{Neutrino decay to scalar component}\label{sec3}
The theoretical model presented in this section represents the exact formulation of the neutrino decay process, while an approximate method may exist in the literature \cite{Curved7.Sadjadi}.
Using (\ref{eqn26}), the evolved neutrino state corresponding to the flavor $\alpha$ at the spacetime point $(t,r)$ can be suggested as follows:
\begin{eqnarray}\label{eqn28}
\ket{\upnu(\theta,r,t)}_\alpha = \sum_{i,\beta} \mathcal{F}_i (r,t) \upnu_i(r_0,t_0) U_{\beta i}(\theta,r) U_{\alpha i}^*(\theta_M , r_0) \ket{\nu_\beta},
\end{eqnarray}
where $U_{\alpha i}(\theta_M,r_0)$ is the ($\alpha i$) element of the mixing matrix in the matter at the production point and $U_{\beta i}(\theta,r)$ is the ($\beta i$) element of the mixing matrix in the vacuum at the detection point.
Assume that at $(t_0,r_0)$, we have a specific flavor neutrino, e.g., $\nu_\alpha$, requires that $\upnu_i(r_0,t_0)=\upnu_0$ for all $i$'s.
Then,
\begin{eqnarray}\label{eqn29}
P_{\alpha \beta} =  \left|\sum_i \mathcal{F}_i(r,t) U_{\beta i}(\theta,r) U_{\alpha i}^*(\theta_M,r_0)\right|^2.
\end{eqnarray}
gives the probability of the flavor transition $\alpha\to \beta$ on the Earth.
For very fast oscillations, the phase parts may be averaged out and, therefore, various probabilities take the forms
\begin{eqnarray}\label{eqn30}
<P_{ee}> = c_{12}^2 c_{13}^2 \left[c_{12}^{M} c_{13}^{M}\right]^2 \mathcal{D}_1 + s_{12}^2 c_{13}^2 \left[s_{12}^{M} c_{13}^{M}\right]^2 \mathcal{D}_2 + s_{13}^2 \left[s_{13}^{M}\right]^2 \mathcal{D}_3,
\end{eqnarray}
\begin{eqnarray}\label{eqn31}
\begin{split}
&<P_{e\mu}> = \left[s_{12}^2 c_{23}^2 + c_{12}^2 s_{23}^2 s_{13}^2 + \frac{1}{2} \sin 2\theta_{12} \sin2\theta_{23} s_{13} c_{\delta}\right] \left[c_{12}^{M} c_{13}^{M}\right]^2 \mathcal{D}_1 \\& ~~~~~~~~~~~ + \left[c_{12}^2 c_{23}^2 + s_{12}^2 s_{23}^2 s_{13}^2 - \frac{1}{2} \sin 2\theta_{12} \sin2\theta_{23} s_{13} c_{\delta}\right] \left[s_{12}^{M} c_{13}^{M}\right]^2 \mathcal{D}_2 \\& ~~~~~~~~~~~ + \left[s_{23}^2 c_{13}^2\right]\left[s_{13}^{M}\right]^2 \mathcal{D}_3,
\end{split}
\end{eqnarray}
and $<P_{e\tau}>$ can be obtained by replacing $c_{23} \rightarrow -s_{23}$ and $s_{23} \rightarrow c_{23}$ in Eq.(\ref{eqn31}), where $c_\delta = \cos\delta_{\text{CP}}$.
We know that all above quantities with superscript $M$ are effective ones in matter.
As a considerable result, it is worth noting that the total probability $<P_{\text{tot.}}>$ (i.e., sum over all probabilities) is not conserved.
In this case, we have $<P_{\text{tot.}}> \leq 1$, where the equality holds if and only if the $\mathcal{D}$-factors are all set to zero.

The probability
\begin{eqnarray}\label{eqn32}
\begin{split}
\delta P_{eX} = 1- <P_{\text{tot.}}>  = 1- \mathcal{D}_{1} \left[c_{12}^{M} c_{13}^{M}\right]^2 - \mathcal{D}_{2} \left[s_{12}^{M} c_{13}^{M}\right]^2 - \mathcal{D}_{3} \left[s_{13}^{M}\right]^2,
\end{split}
\end{eqnarray}
can calculate the difference in the total probability from unity.
From KamLAND \cite{KamLAND}, SNO\cite{SNO}, and SK\cite{Super-Kamiokande} experiments, it is accepted that the mixing between various eigenstates of neutrinos is the most significant reason for the solar neutrino problem.
However, all the probabilities above suffer explicitly from damping/enhancing factors denoting an extra deficit in the number of neutrinos on the Earth.
This deficit can be considered a result of a phenomenon called neutrino decay.
The subscript $X$ refers to the particle corresponding to the dark energy scalar boson to which neutrinos can decay through the non-standard conformal coupling.
As in Ref.\cite{MohseniSadjadi-Anari1}, such an NSI might modify the neutrino and scalar field density continuity equations, which results in a possible alleviation of the coincidence problem and explains the ignition of the present acceleration of the Universe in the non-relativistic era of mass varying neutrinos (at redshift $z \sim 10$) \cite{Fardon}.

As a particular case of interest, suppose that only the lightest neutrino mass eigenstate, $\nu_1$, is stable during neutrino propagation, i.e., $\mathcal{D}_{1} = 1$, and the other two mass states, $\nu_2$ and $\nu_3$, are coupled to the scalar field.
The decay probability of the $\nu_e \rightarrow \phi$-scalar field is therefore given by
\begin{eqnarray}\label{eqn33}
\delta P_{e\phi} =  1 - \left[c_{12}^{M} c_{13}^{M}\right]^2 - \mathcal{D}_{2} \left[s_{12}^{M} c_{13}^{M}\right]^2 - \mathcal{D}_{3} \left[s_{13}^{M}\right]^2.
\end{eqnarray}
In such a case, WEP is violated ($\beta_1 = 0$, whereas $\beta_2,\beta_3 \ne 0$).
In dense regions where the scalar field is highly screened, WEP violation may not be detected by local gravitational tests, but there is always the possibility that violation happens at the large scales in the Chameleon screening model \cite{Waterhouse,Khoury-Weltman,Burrage}.
Furthermore, for vanishing smallest leptonic mixing angle $\theta_{13}^M$, the contribution of $\nu_3$-$\phi$ coupling will disappear, i.e., we have
\begin{eqnarray}\label{eqn34}
\delta P_{e\phi} = \left(1-\mathcal{D}_2\right) \left[s_{12}^{M}\right]^2,
\end{eqnarray}
which increases when $\mathcal{D}_2$-factor declines.
Experimental constraints on invisible neutrino decay come mainly from the solar and reactor neutrino experiments.
A study has been done to look for non-radiative neutrino decay.
The strongest constraint on the lifetime of $\nu_2$ is $\tau_2/m_2 > 1.92 \times 10^{-3}$ sec./eV \cite{SNO-Neut.Decay}.

In addition, the damped two-flavor electron-neutrino transition probabilities on the Earth are given by
\begin{eqnarray}\label{eqn35}
	<P_{ee}> = \cos^2\theta \cos^2\theta_{M} \left(\mathcal{D}_1\right) + \sin^2\theta \sin^2\theta_{M} \left(\mathcal{D}_2\right),
\end{eqnarray}
and
\begin{eqnarray}\label{eqn36}
	<P_{e\mu}> = \sin^2\theta \cos^2\theta_{M} \left(\mathcal{D}_1\right) + \cos^2\theta \sin^2\theta_{M} \left(\mathcal{D}_2\right).
\end{eqnarray}
There are two interesting cases, differing in coupling parameter $\beta$:
\paragraph{Case 1} First, we respect the WEP if we assume that $\mathcal{D}_1 = \mathcal{D}_2 = \mathcal{D}$.
We have
\begin{eqnarray}\label{eqn37}
	\begin{split}
		&<P_{ee}> = \frac{\mathcal{D}}{2} \left[1 + \cos2\theta \cos 2\theta_M\right],\\&
		<P_{e\mu}> = \frac{\mathcal{D}}{2} \left[1 - \cos2\theta \cos 2\theta_M\right],
	\end{split}
\end{eqnarray}
which are the common probabilities multiplied by a $\mathcal{D}$-factor \cite{Gonzalez-Garcia-Nir}.
\paragraph{Case 2} Second, if $\mathcal{D}_1 = 1$, but $\mathcal{D}_2  \ne 1$.
Then, defining $\mathcal{D}_2 = \mathcal{D}$, we have
\begin{eqnarray}\label{eqn38}
	\begin{split}
		&<P_{ee}> = \cos^2\theta \cos^2\theta_{M} + \sin^2\theta \sin^2\theta_{M} \left(\mathcal{D}\right), \\&
		<P_{e\mu}> = \sin^2\theta \cos^2\theta_{M} + \cos^2\theta \sin^2\theta_{M} \left(\mathcal{D}\right).
	\end{split}
\end{eqnarray}
This case is devoted to the decay process of the heavier mass eigenstate ($\nu_2$) \cite{Curved7.Sadjadi}, which means that invisible neutrino decay leads to the exponential disappearance of the heavy component of the propagating neutrino.
As mentioned before, we note that WEP may be violated in such a case.

\subsection{Effective mixing parameters}
Diagonalizing the three-flavor neutrino evolution Hamiltonian in equation (\ref{eqn27}) is rather complicated and cumbersome. Hence, we consider a simplified two-flavor case inside matter.
Using Eq.(\ref{eqn27}), the Hamiltonian which governs the propagation of neutrinos inside matter is obtained by
\begin{eqnarray}\label{eqn39}
\mathcal{H} = \frac{1}{2E_\nu} \left[ U
\begin{pmatrix}
m_1^{\prime 2} & 0\\
0 & m_2^{\prime 2}
\end{pmatrix}
U^\dagger + 2E_\nu V^\prime(r)
\begin{pmatrix}
1&0\\
0&0
\end{pmatrix}
\right].
\end{eqnarray}
Hereafter, the critical point is to diagonalize the effective Hamiltonian (\ref{eqn39}) and the derivation of the explicit expressions for the effective oscillation parameters.
The solar neutrino parameters of this model correspond to the LMA-MSW solution.
It should be recalled that KamLAND \cite{KamLAND} has measured the solar mixing angle ($\theta_{12}$) in almost vacuum to be entirely consistent with that from solar experiments such as SNO \cite{SNO} and SK \cite{Super-Kamiokande}.

After doing some simple calculations, the Hamiltonian above is given by
\begin{eqnarray}\label{eqn40}
\mathcal{H} = \frac{1}{4E_\nu} \left[\Delta m^{\prime 2}
\begin{pmatrix}
-\cos 2\theta&\sin 2\theta \\
\sin 2\theta & \cos 2\theta
\end{pmatrix}
+ 2E_\nu V^\prime(r)
\begin{pmatrix}
1&0\\
0&-1
\end{pmatrix}
\right].
\end{eqnarray}
Then the effective neutrino oscillation parameters when taking into account matter NSIs are
\begin{eqnarray}\label{eqn41}
\Delta m^2_M = \sqrt{\left[\Delta m^{\prime 2} \sin 2\theta\right]^2 + \left[\Delta m^{\prime 2} \cos 2 \theta - 2E_\nu V^\prime(r)\right]^2},
\end{eqnarray}
and
\begin{eqnarray}\label{eqn42}
\sin 2\theta_{M} = \frac{\Delta m^{\prime 2} \sin 2\theta}{\Delta m^2_M}.
\end{eqnarray}
These effective oscillation parameters depend on the scalar field, neutrino energy, and of course on the electron number density of the Sun.
From now on, it is more convenient to consider neutrinos would exactly behave as they do in the vacuum except possessing new mass and mixing parameters \cite{Zhang}.

For neutrinos produced at the center of the Sun, we also consider a phenomenon corresponding to the maximal mixing inside matter, i.e., $\sin 2\theta_M =1$.
The quantity
\begin{eqnarray}\label{eqn43}
\frac{\Delta m^{\prime 2}(r_{\text{res.}}) \cos 2 \theta}{2\sqrt{2}G^\prime_F(r_{\text{res.}}) N_{e}(r_\text{res.})} = E^{\text{res.}}_{\nu}
\end{eqnarray}
gives the resonance energy.
Since the resonance energy is explicitly dependent on the neutrino-scalar coupling functions $A(\phi)$, there would be a shift in the resonance energy for various coupling strengths.

\section{Results and discussions}\label{sec4}
Clear evidence that neutrinos may oscillate in various flavors comes from diverse experiments, which indicate that physics beyond the standard model is required to properly deal with the neutrino properties.
An important related aspect is their interaction with external fields, e.g., electromagnetic or gravitational fields \cite{Giunti,Piriz}.
The formulas of oscillation probabilities of neutrinos with different flavors might be affected by those external fields, compared to the formulas calculated for the vacuum case.
For instance, the interactions of neutrinos with external electromagnetic fields may induce the so-called spin oscillation and/or spin-flavor oscillations \cite{Giunti}.
It is also well-known that the propagation of neutrinos in the curved spacetime may affect the neutrino oscillation probability \cite{Curved.Cardall,Curved.Fornengo,Curved2.Visinelli,Curved4,Curved5,Curved6,Curved7.Sadjadi,Curved8.Yaz}, or induce the change of the polarization of a spinning particle \cite{Obukhov}.
In a similar way, we have studied the effects of non-standard neutrino interactions with the Chameleon scalar field, which governs the neutrino decay process.

Performing the numerical computations in all formulas, one arrives at the results depicted hereafter.
Throughout our analysis, we will take the numerical values $\tan^2 \theta_{12}=0.41$, $\sin^2 2\theta_{23} \simeq 0.99$, $\sin^22\theta_{13} \simeq 0.09$,
\[ \Delta m_{21}^2=7.4 \times 10^{-5} \text{eV}^2, \] and \[|\Delta m_{23}^2|=2.5 \times 10^{-3} \text{eV}^2\]
for neutrino mass and mixing parameters \cite{Esteban:2020cvm}, and $n=1$, and $M \simeq 2.08~\text{keV}$ for Chameleon scalar field constrained by cosmological observations \cite{Waterhouse}. 
Nevertheless, we expect the Borexino experiment \cite{Agostini} to put a constraint on the mass-scale parameter $M$, which will be in a good agreement with the value mentioned above (see Fig.\ref{fig6}).

\begin{figure}[H]
	\centering
	\includegraphics[scale=0.45]{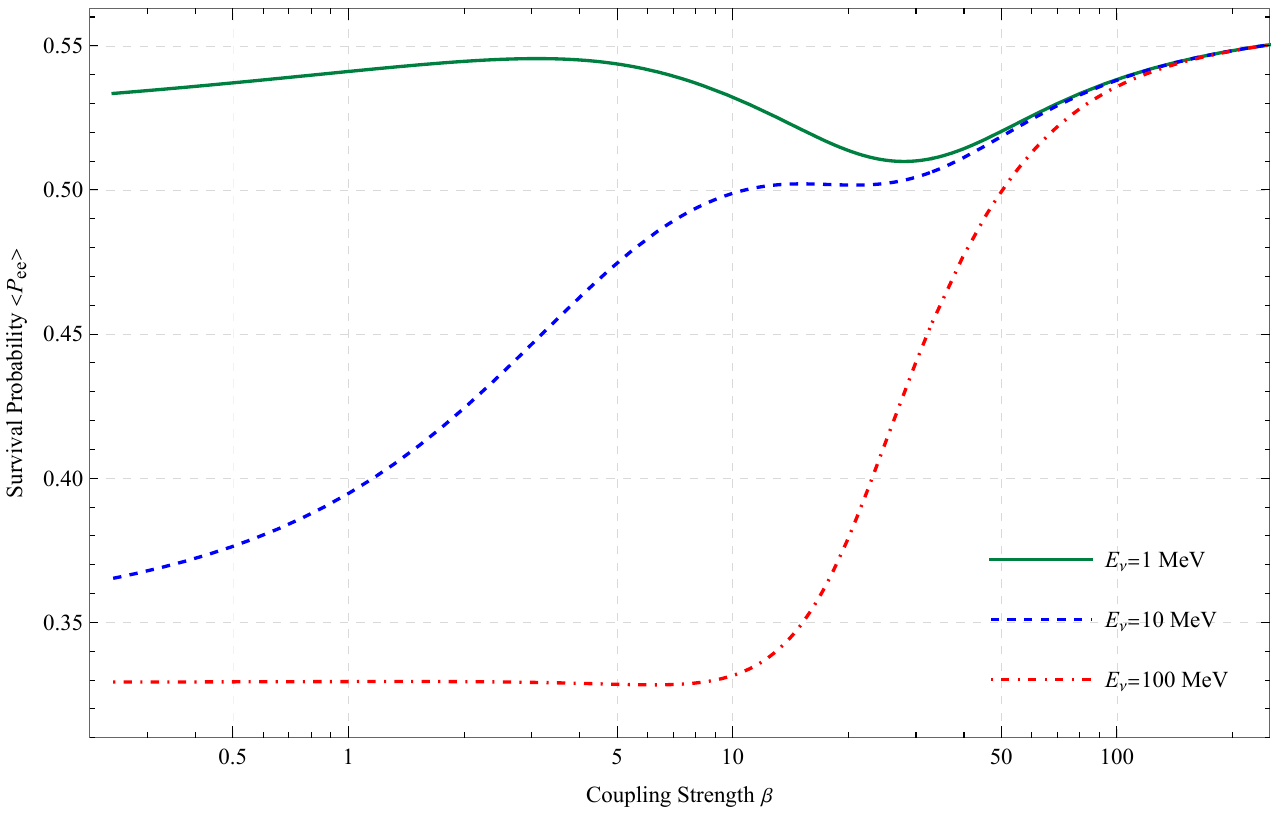}
	\caption{
		\footnotesize{This figure shows that the survival probability in terms of the neutrino-scalar coupling parameter $\beta$ for three values of neutrino energy.
		All curves obtain the same values at the $\beta \gtrsim 100$.}}
	\label{fig1}
\end{figure}

In the standard framework, we know that the neutrino oscillation probabilities are functions of eight parameters [$L$ (the experiment baseline), $E_\nu$; $\theta_{12}$, $\theta_{13}$, $\theta_{23}$, $\delta_{\text{CP}}$, $\Delta m^2_{21}$, and $\Delta m^2_{31}$], whereas for matter NSIs, they will be depending on one additional parameter $\beta$ (see equations (\ref{eqn30}) and (\ref{eqn31})).
Analytical expressions for the neutrino oscillation probabilities in the presence of the matter NSIs are derived as an extra modifications, i.e., neutrino decay, to the solar neutrino problem, see Eq.(\ref{eqn32}).
Thus, in figure \ref{fig1}, we show the $\nu_e \to \nu_e$ survival probability as the NSI parameter $\beta$ is varied. 
The solid green curve corresponds to the survival probability when the neutrino energy is considered to be $E_\nu = 1$ MeV for $^7$Be and $pep$ neutrinos.
Whereas, for $^8$B neutrinos (with $E_\nu = 10$ MeV) the survival probability in terms of the coupling parameter obtains less values for $\beta \lesssim 100$.
According to the figure, the existing effects on $<P_{ee}>$ in terms of neutrino energy are stronger for $\beta \lesssim 100$, resulting in observable differences in the flux of solar electron-neutrinos and it will also affect the measurement of parameters.
At high neutrino energies ($E_\nu = 100$ MeV), the survival probability has remained constant at $<P_{ee}> \sim 0.33$ for lower coupling parameters.

Then, all the curves are exactly overlapping for values of NSI coupling parameter $\beta \gtrsim 100$.
From this observation, one would see that for this range of the NSI parameter the sensitivity of $\nu_e$ survival probability to the neutrino energy is completely lost.
In other words, if the NSI parameter could be considered $\beta \gtrsim 100$, then it would be possible to detect neutrinos with energies even below Borexino threshold ($\sim 0.19$ MeV) \cite{Borexino}.
But, we know that the background of Borexino experiment is dominated by irreducible neutrino components basically due to the radioactive isotopes
contaminating the scintillator itself. 
At low-energies, for instance, ($^{14} \text{C}$) $\beta^-$ decay process ($Q = 0.156$ MeV) is the main background for $pp$ neutrinos, while for two mono-energetic $^7$Be solar neutrino lines, the $\beta^-$ decays of $^{85}$Kr ($Q = 0.687$ MeV) and $^{210}$Bi ($Q=1.160$ MeV), and $\alpha$ decay of $^{210}$Po ($E_\alpha = 5.3$ MeV) are considered the main backgrounds for the detection of electron recoil spectra \cite{Agarwalla}.
All these imply that new physics can only be probed down to $E_\nu \sim 10~\text{MeV}$ so that the NSI effects are stronger at $^8$B neutrino energies but not particularly large at $^7$Be neutrino energies, limiting the sensitivity of Borexino to the scalar-neutrino coupling $\beta$.
As will be shown, comparing the SNO experimental data \cite{SNO-Neut.Decay} to the present model illustrates that the coupling parameter $\beta$ is restricted to the values lower than $100$ (see the discussion around Fig.\ref{fig3}).
Furthermore, the modified matter effects in the survival probability $<P_{ee}>$ are negligible and the vacuum oscillation is dominated for large coupling strengths, i.e., for $\beta \gtrsim 100$, in comparison with data imposed by Borexino.
So, this inconsistency of model (at couplings $\beta \gtrsim 100$) with observational data constrains the coupling parameter to lower values, i.e., to $\beta < 100$ (see the discussion around Fig.\ref{fig4}) such that large values of $\beta$ does not need to be included.

\begin{figure}[H]
	\centering
	\includegraphics[scale=0.46]{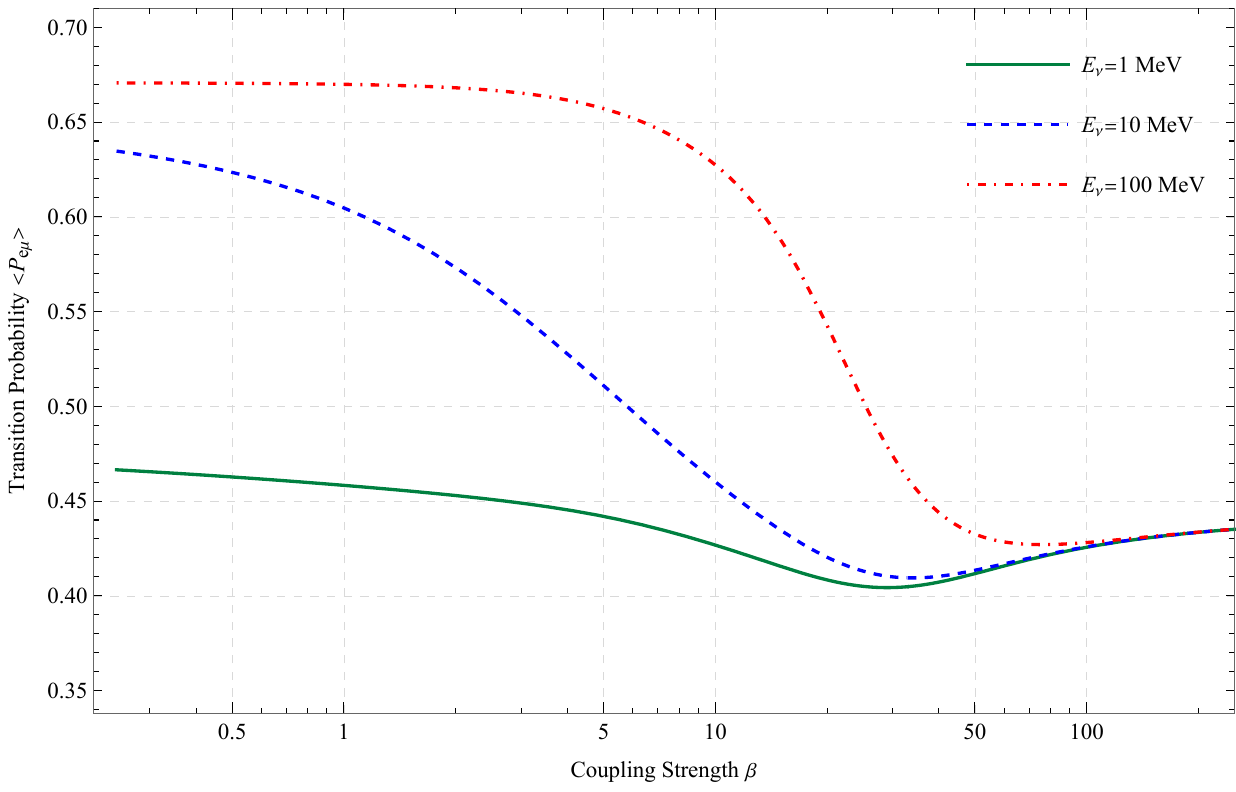}
	\caption{\footnotesize{Transition probability in terms of the coupling parameter for different values of neutrino energy.
		For $\beta \gtrsim 100$, the probability $<P_{e\mu}>$ acquires the same value for all energies.}}
	\label{fig2}
\end{figure}

\begin{figure}[H]
	\begin{subfigure}{.5\textwidth}
		\centering
		\includegraphics[scale=0.61]{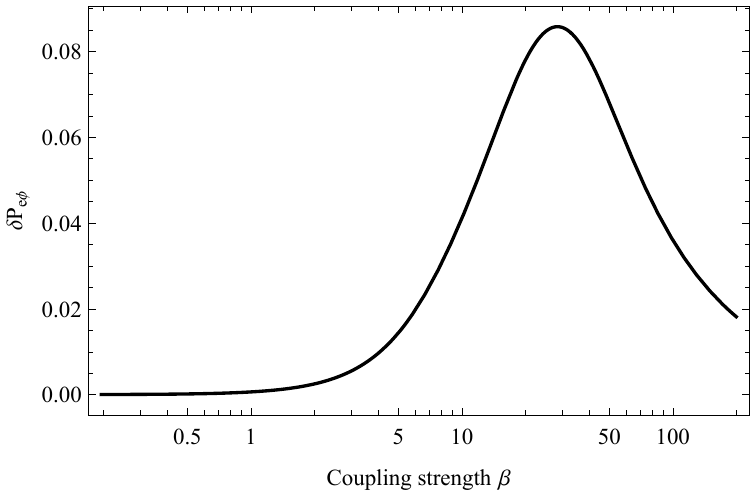}
		\label{fig3-1}
		\caption{}
	\end{subfigure}
	\begin{subfigure}{.5\textwidth}
		\centering
		\includegraphics[scale=0.42]{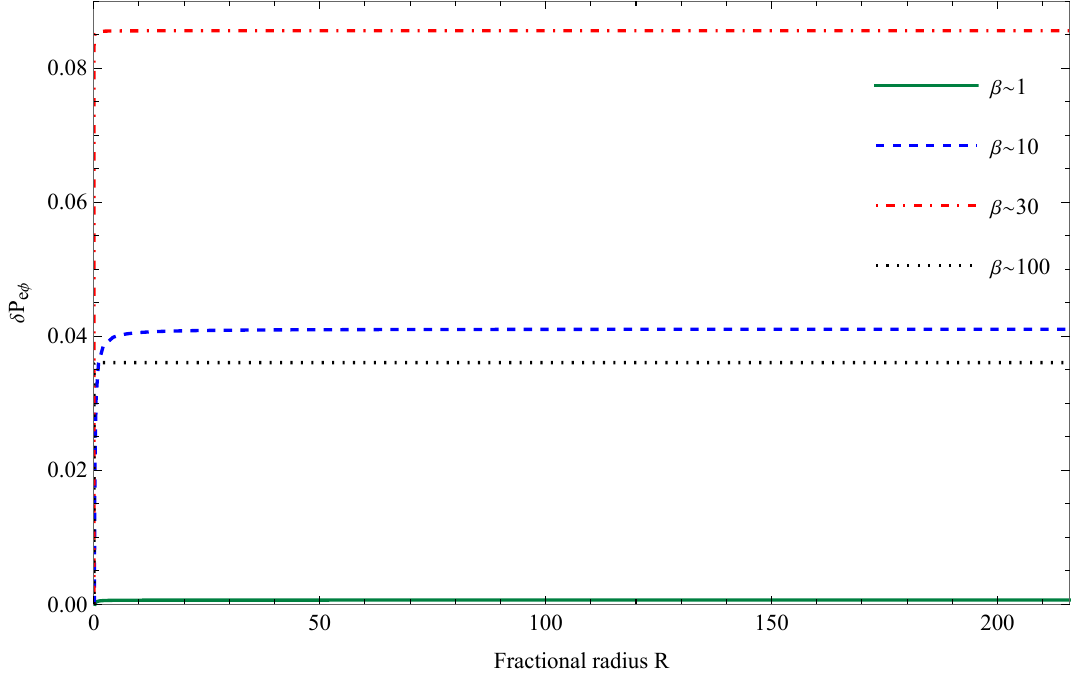}
		\label{fig3-2}
		\caption{}
	\end{subfigure}
	\caption{\footnotesize{The loss (or decay) probability of $^8$B electron-neutrinos with energy $E_\nu = 10$ MeV on the Earth (a) with respect to the coupling parameter $\beta$, and (b) with respect to the fractional radius $R$.
		As can be seen, the maximum value belongs to the case with $\beta \simeq 30$. We note that $\delta P_{e\phi} (\text{on Earth}) \simeq 0.001$ for $\beta \sim 1$.}}
	\label{fig3}
\end{figure}

\begin{figure}[H]
	\centering
	\includegraphics[scale=0.43]{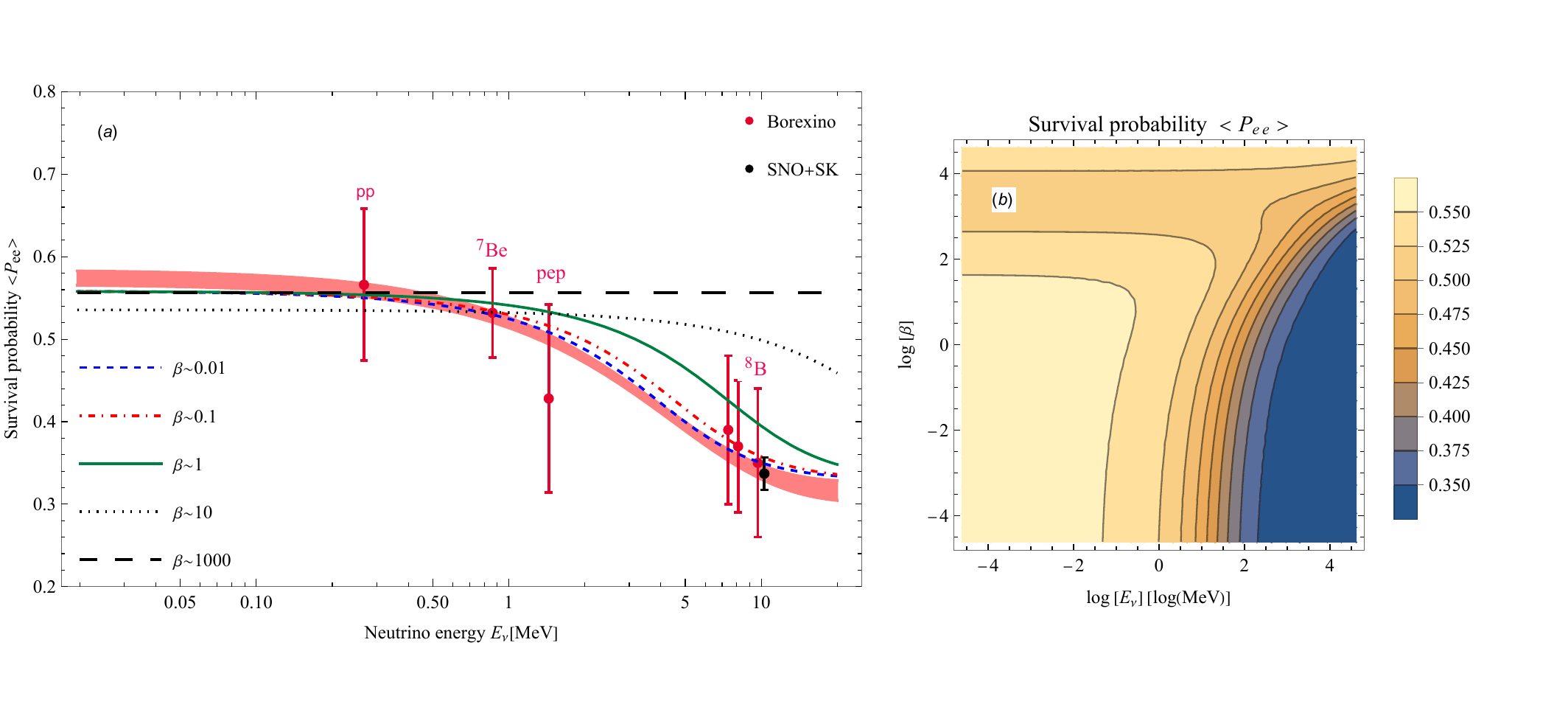}
	\caption{\footnotesize{(a) Electron-neutrino survival probability $<P_{ee}>$ as a function of its energy for LMA-MSW case with only SI effects taken into account (pink band), and LMA-MSW + neutrino-scalar NSI cases for $\beta \in \{0.01,0.1,1,10,1000\}$.
		The observational data are taken from Borexino \cite{Agostini} and SNO+SK \cite{Zyla-PDG}.
	(b) The electron-neutrino survival probability $<P_{ee}>$ as a function of coupling parameter and neutrino energy. The colored regions correspond to the numerical results (darker regions for lower probabilities).
}}
	\label{fig4}
\end{figure}

For comparison, we also show the transition probability in terms of the NSI parameter $\beta$ in Fig.\ref{fig2}.
Increasing the NSI coupling parameter, the effects become different for various neutrino energies.
As shown in Fig. \ref{fig2}, the variation is large for higher-energy neutrinos in the range $0 \lesssim \beta \lesssim 100$, while $^7$Be and $pep$ neutrinos do not experience intense changes.
For larger coupling strengths, i.e., for $\beta \gtrsim 100$, however, we expect that the probability of conversion $\nu_e \to \nu_\mu$ loses its sensitivity to neutrino energy.

The electron-neutrino survival probability including damping factor has been given in Eqs.(\ref{eqn30}) and (\ref{eqn31}). 
The interesting point in the present model is that the sum over all probabilities is not equal to unity.
The missing value of the total conversion probability can be interpreted as a decay to yet unknown states.
Although neutrino decay is now not known as the dominant effect behind the solar neutrino problem \cite{SNO-Decay1,SNO-Decay2}, solar neutrinos make an opportunity for probing decay possibility as a second-order effect.
There are generally two kinds of decays depending on whether the final state particles are only \textit{invisible}, such as sterile neutrinos (generally non-active neutrino flavors), Majorons, etc., or they include \textit{visible} particles too, e.g., active neutrino flavors \cite{Baerwald}.

Invisible decays refer to the scenarios where the decay products do not interact with the detector's medium. 
This is the case, for example, when the neutrinos decay into new exotic particles with no (or very suppressed) interactions (following the screening mechanism) or into standard particles with energies below detection threshold.
The former case, i.e., neutrino decay to an exotic field, can be interpreted as neutrino decay to $\phi$-particles, which are responsible for the accelerated expansion of the Universe today.

To clarify our model, we take a numerical example for $\delta P_{e\phi}$ in figure \ref{fig3}.
In the case of exact expression for the ``loss probability'' shown in the left plot of Fig.\ref{fig3}, we observe that the loss probability $\delta P_{e\phi} \simeq 0.086$ (on Earth) is maximal for $\beta \sim 30$, while $\delta P_{e\phi} \lesssim 0.001$ for $\beta \lesssim 1$.
In the right panel, we have plotted the variation of total probability with respect to the distance from the source, in which the discrepancy
grows to a constant value at the surface of the Sun and remains constant outside the body to the Earth.
So, a nonzero loss probability leads to an exponential decay of neutrinos relative to the $\beta \to 0$ limit and modifies the amount of electron-neutrinos reaching the detector.

Also, comparing the $\mathcal{D}$-factor of the present model (Eq.(\ref{eqn24})) with $\beta_1 = 0$ and $\beta_2 \ne 0$ (\textbf{Case 2} in section \ref{sec3}) to the common decay factor \cite{SNO-Neut.Decay}, i.e., $\exp \left[- \frac{L}{E_\nu} . \frac{m_2}{\tau_2}\right]$, might restrict the values of coupling parameter $\beta_2$. 
Accordingly, the solar neutrino lifetime is then given by
\begin{eqnarray}\label{eqn44}
\frac{\tau_2}{m_2} = \frac{2L M_p}{3 \beta_2 E_\nu \left[\phi(L) - \phi(L_0)\right]},
\end{eqnarray}
where $t\simeq L$ (for ultra-relativistic neutrinos) is the Earth-Sun distance ($t \simeq 500$ sec.).
Fitting the combined SNO + other solar neutrino experiments \cite{SNO-Neut.Decay} yields $\frac{\tau_2}{m_2} > 1.92 \times 10^{-3}$ sec./eV at $99 \%$ C.L., corresponding to $\beta_2 \lesssim 100$.
These requirements restrict the effective mass of the Chameleon scalar field deep in space (in the dilute regions) $m_{\text{eff.}}(\beta_2) < 1.12 \times 10^{-29}$ eV.
However, for the coupling parameter $\beta_2 \simeq 30$ in which the loss probability (or the deficit in the neutrino flux caused by decay process) is maximal, the neutrino lifetime equals to $\frac{\tau_2}{m_2} \simeq 8.62 \times 10^{-3}$ sec./eV.

Each of the probabilities shown are calculated numerically, using $E_\nu = 10$ MeV for neutrino energy and $\rho_0 \sim 10^{-24}$ g.cm$^{-3}$ for background matter density. 
Out of the two screening mechanisms, i.e., the Chameleon and Symmetron, the decay process $\nu \to \phi$ due to the Chameleon is stronger (see Ref.\cite{Curved7.Sadjadi}).
This difference may has roots in their different coupling strengths to neutrinos.

While many experiments are programmed to detect matter-affected neutrino oscillations (the MSW effect) on the Earth and look for extra effects beyond standard model, the only significant matter effect observed in experiments is for solar neutrinos. 
Thus, solar neutrinos might be used to see the non-standard effects on the matter flavor oscillations. 
Different probabilities are particularly sensitive to models of NSI of neutrinos with a scalar field, leading to the neutrino decay and to put bounds on the NSI parameter.

In Fig.\ref{fig4}(a), we show the predictions of survival probability $<P_{ee}>$ for solar neutrinos and for several representative values of NSI parameter $\beta$, spanning the energy regime from the lowest-energy $pp$ to the higher-energy $^8$B neutrinos.
The predicted survival probability might be in three regimes.
At low energies (for $pp$ neutrinos), vacuum effects are dominant, and thus the survival probability corresponds to the vacuum value (about $0.55$). 
Between about $1$ MeV and $4$ MeV, the survival probability decreases from the vacuum to the matter-dominated value (about $0.33$). 
It is in this transition region where the non-standard effects would be most considerable, as they have intersections with the curve from standard matter effects (light-pink band).
The light-pink band is the best theoretical prediction of $<P_{ee}>$ (within $\pm 1 \sigma$) according to standard MSW-LMA solution. 
We guess that the best fit to this curve can be given by
\begin{eqnarray}\label{eqn45}
	<P_{ee}(E_\nu)> \simeq 0.322 + 0.244~e^{-0.25(\frac{E_\nu}{\text{MeV}})}.
\end{eqnarray}
This behavior might be better described in a contour-plot of Fig.\ref{fig4}(b), which shows the electron-neutrino survival probability as a function of the coupling parameter $\beta$ and neutrino energy $E_\nu$.

In what follows, a general discussion about the survival probability for different ranges of the coupling parameter $\beta$ has been given:

$\bullet$ For $\beta < 0.01$, the effects of the NSI modifications will be eliminated and, olny the standard MSW effect will be left.
So, setting $\beta \to 0$ signifies the no-Chameleon case.

$\bullet$ The theoretically predicted shift of $<P_{ee}>$ is mostly within the error bars of the experimentally determined values of Borexino for $0.01 \lesssim \beta \lesssim 5$.
In the Chameleon model proposed by Khoury and Weltman \cite{Khoury-Weltman}, $\beta \sim 1$ implies the gravitational strength for the fifth force mediated by the scalar field.

$\bullet$ In region $5 < \beta \lesssim 30$, increasing $\beta$ will decrease the amplitude of the survival probability because of the damping signatures, such that the minimum will happen at $\beta \simeq 30$ (as expected from figure \ref{fig3}).

$\bullet$ Then for $30 <\beta \lesssim 100$, the NSI effects on the MSW effect push the curves to the vacuum survival probability.
Finally, as the satellite experiments \cite{SatExp1,SatExp2,SatExp3} have proposed, they are unable to put an upper bound on the coupling parameter \cite{Mota}, so we may consider cases with $\beta \gtrsim 100$ even for $\beta \to 1000$, in which $<P_{ee}>$ implies the vacuum-LMA scenario (to the constant value $\sim 0.55$) that assumes all solar neutrinos oscillate in the vacuum regime independent of neutrino energy.
By comparing to the Borexino data, however, we can restrict the scalar-neutrino coupling $\beta$.
The modified matter effects are stronger for $^8$B neutrino energies such that the survival probability is negligible and the vacuum oscillation is dominated for large coupling strengths, i.e., for $\beta \gtrsim 100$.
So, this inconsistency of model with data constrains the coupling parameter to lower values $\beta < 100$.

As the electron number density $N_e(r)$ decreases towards the surface of the Sun, the resonance energy $E^\text{res.}_\nu$ will increase (see Eq.(\ref{eqn43})).
The presence of non-zero coupling parameter $\beta$ will also shift the resonance energy such that stronger coupling $\beta$ results in a higher value of $E^{\text{res.}}_\nu$.

\begin{figure}[H]
	\centering
	\includegraphics[scale=0.43]{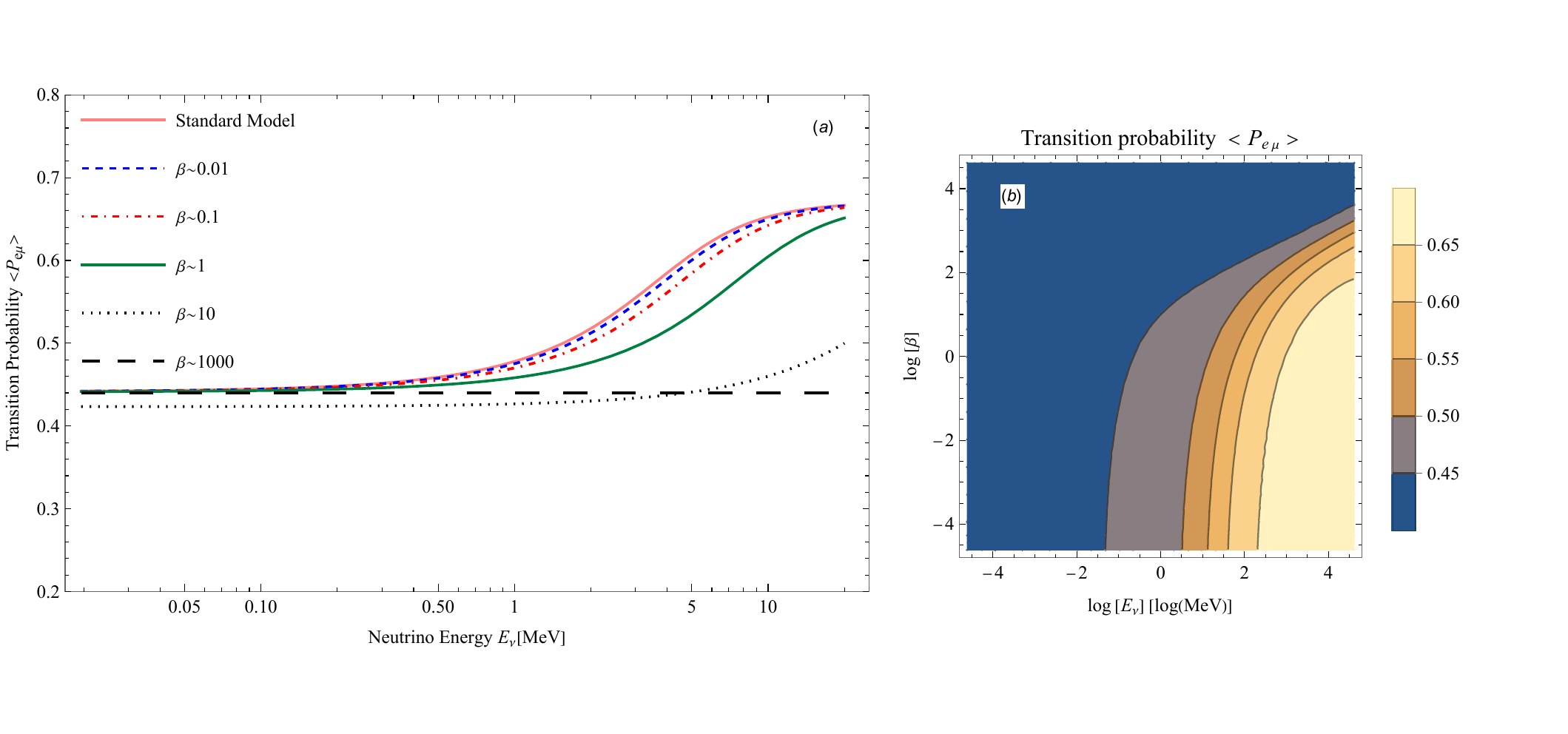}
	\caption{\footnotesize{(a) The $\nu_e \to\nu_\mu$ transition probability with respect to the $E_\nu$. The figure is plotted for both SI and NSI with different coupling strengths.
(b) We present the contour plot of neutrino transition probability $P_{e\mu}$ for different values of the coupling parameter and neutrino energy.
}}
	\label{fig5}
\end{figure}

\begin{figure}[H]
	\centering
	\includegraphics[scale=0.43]{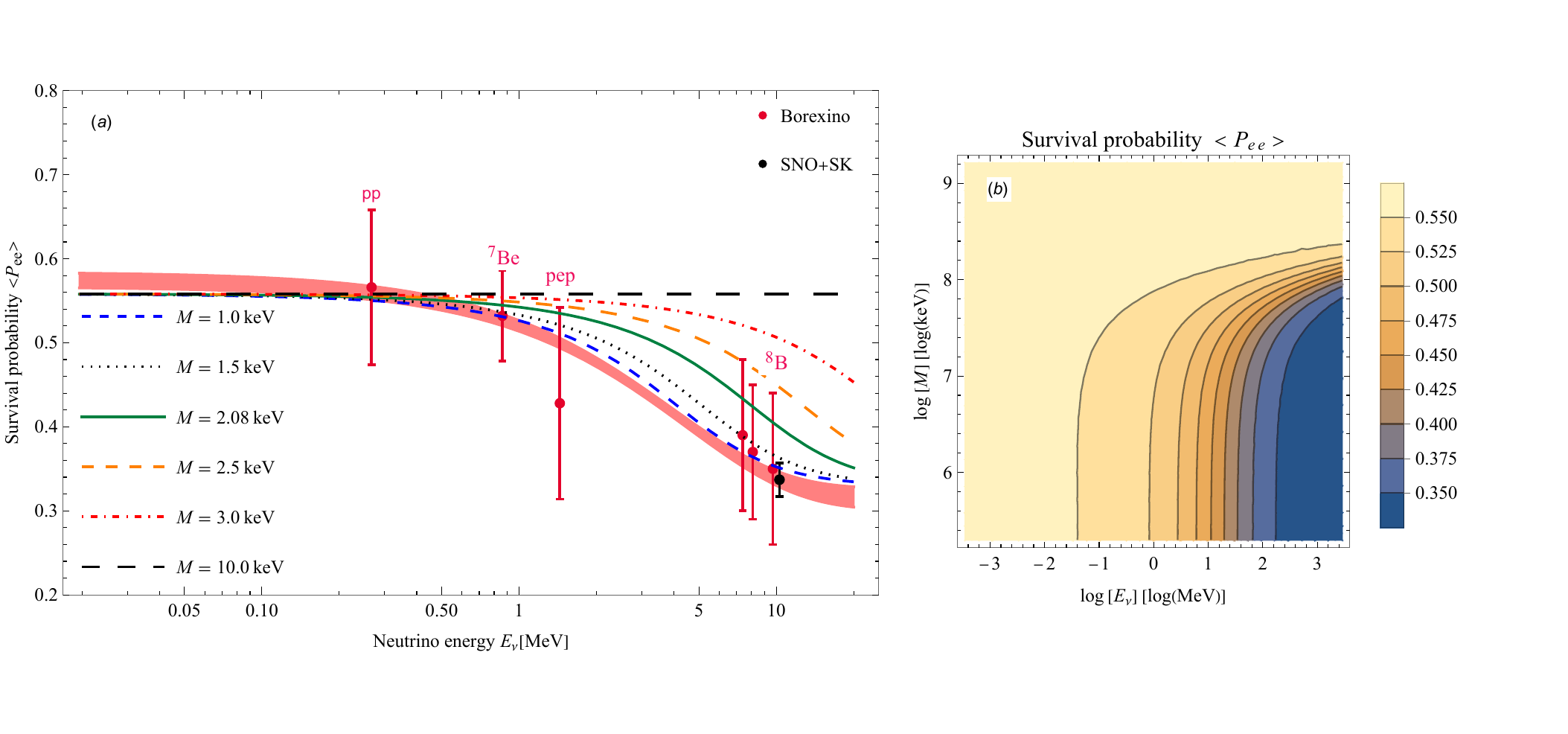}
	\caption{\footnotesize{(a)  Electron neutrino survival probability as function of the neutrino energy for various mass-scale parameters $M$ and for $\beta \sim 1$, evaluated for the $^8$B neutrino source assuming the BS05 standard solar model \cite{Bahcall-Density}. Points represent the results from $pp$ to $^8$B measurements of Borexino \cite{Agostini} and SNO+SK \cite{Zyla-PDG}.
			(b) The electron-neutrino survival probability $<P_{ee}>$ as a function of mass-scale parameter $M$ and neutrino energy. The colored regions correspond to the numerical results.
	}}
	\label{fig6}
\end{figure}

Averaging oscillations over neutrinos with different coupling strengths give smooth transition probabilities too, which is depicted in Fig.\ref{fig5}(a).
In this figure, various curves denote the exact numerical results of the effects of nonzero NSI parameters, and the pink curve represents the probability without including non-standard coupling.
The contour plot of the transition probability with respect to the coupling parameter and neutrino energy has been drawn in Fig.\ref{fig5}(b), preparing a clearer continuous representation of the $\beta$-$E_\nu$ dependence of transition probability.

We have also re-examined the modified MSW effects in our theoretical framework for different mass-scale parameters $M$ and suggested an allowed range for this parameter corresponding to the Borexino observational data \cite{Agostini}. 
As shown in Fig. \ref{fig6}, the allowed range for mass-scale parameter is $1.0~\text{keV}\lesssim M \lesssim 2.5~\text{keV}$, which is also fixed to $M=2.08$ keV from the dark energy calculations \cite{Waterhouse}.
For this range, one can see that the curves of survival probabilities from the present model are well within the error bars of Borexino, while the modified matter effect potential vanishes and the curves tend to vacuum oscillation scenario for $M>2.5~\text{keV}$ at all energies. For the latter case, $\nu_e$-survival probability is given by $<P_{ee}> = 1 - \frac{1}{2} \sin^2(2\theta_{12})\sim 0.55$.

\begin{figure}[H]
	\centering
	\includegraphics[scale=0.56]{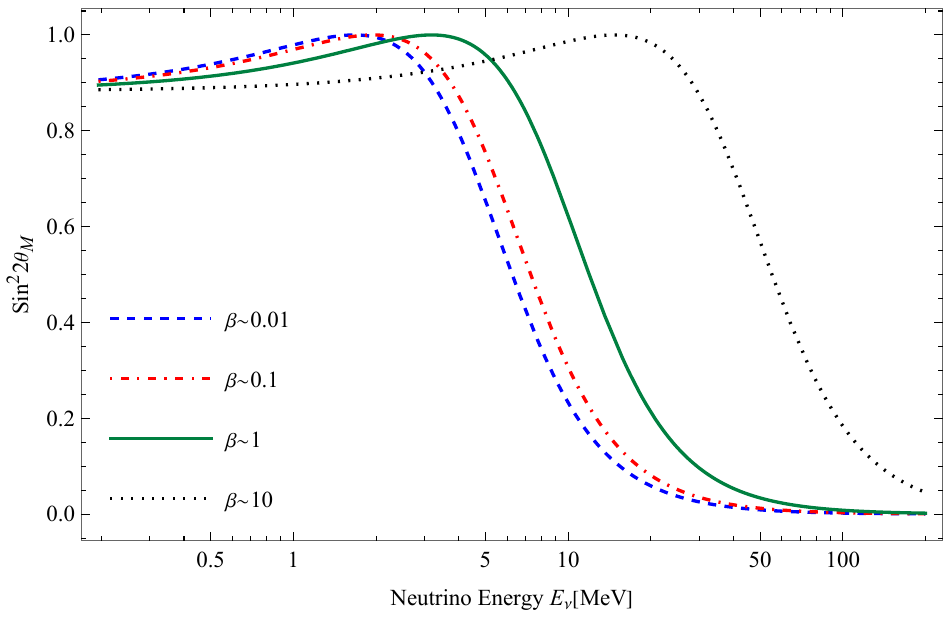}
	\caption{\footnotesize{$\sin^2 2\theta_M$ as a function of neutrino energy $E_\nu$.}}
	\label{fig7}
\end{figure}

According to (\ref{eqn42}), the effective mixing parameter in matter, $\sin^2 2\theta_M$, depends on the electron number density and neutrino energy through the ratio (\ref{eqn43}).
The dependence $\sin^2 2\theta_M (E_\nu)$ for different values of the NSI coupling parameter $\beta$, corresponding to the parameters discussed above, is shown in figure \ref{fig7}.
As shown, increasing $\beta$ results in the resonance happening at higher energies.
At energies less than $\sim 20$ MeV, the standard maximal mixing occurs at the peaks, i.e., $\sin^2 2\theta_M \to 1$.
For higher-energy neutrinos, the mixing inside matter goes to zero; thus, the transition probability becomes negligible in this energy level.

\section{Conclusion}\label{sec5}
Neutrino flavor conversion inside the Sun can be affected via non-standard processes, using the notion of damping behavior through a conformal coupling to the exotic scalar fields.
This phenomenon might mathematically be resulted by new scalar-dependent forms of the mass, wavefunction, and the weak Fermi constant (see (\ref{eqn12})-(\ref{eqn14})), imposing a damping factor in the neutrino flux (or equivalently, the survival probability) implying neutrino decay to unknown states as the sub-leading explanation of the solar neutrino deficit.
This process shows itself explicitly as an exponential $\mathcal{D}$-factor in different exact conversion probabilities obtained for the general 3$\nu$ case (see (\ref{eqn31}) and (\ref{eqn32})).
Moreover, the mass and mixing parameters are affected by the interactions with both matter electrons and scalar field, and can be obtained by diagonalizing the effective Hamiltonian matrix (see (\ref{eqn41}) and (\ref{eqn42})).
As a noteworthy result, sum over all probabilities is not equal to unity caused by the aforementioned $\mathcal{D}$-factor, leading to the loss probability $\delta P_{e\phi}$ (see (\ref{eqn33})).
In addition to the scalar-dependence of the probability amplitude, however, we might focus on the effect of spacetime curvature on the results, which is presented in the phase of oscillations (see (\ref{eqn25})).

An extra aspect existing in this work is the sensitivity available for such effects to be observed in the experiments.
When analyzing neutrino observational data, the usual emphasis is on the survival probability and flux dependence on the neutrino energy; hence, the non-standard effects on the survival probability are strong at $^8$B neutrino energies but not particularly large at $^7$Be neutrino energies, limiting the sensitivity of $<P_{ee}>$ to NSI parameters (see the discussions around figures \ref{fig3}, \ref{fig4}, and \ref{fig6}).
For large coupling strengths, i.e., for $\beta \gtrsim 100$, the electron-neutrino survival probability loses its dependence to the neutrino energy such that neutrinos with energies below Borexino threshold will possibly be detected.
However, the non-standard couplings of neutrinos may not possibly show their influences at low energies (from $pp$ to $^7$Be neutrinos) in the Borexino experiment, because Borexino's background is dominated by irreducible neutrino components mainly caused by the radioactive isotopes contaminating the scintillator's medium at low energies.
This may constrain the NSI parameter to be $\beta \lesssim 100$ (see the discussions around figure \ref{fig1}).
Moreover, fitted solar neutrino data of the neutrino lifetime constrain the range of the coupling parameter to be $\beta \lesssim 100$ (see the discussion below (\ref{eqn44})), so very large values of $\beta$ does not need to be included.
As a numerical result, the neutrino lifetime is $\frac{\tau_2}{m_2} \simeq 8.62 \times 10^{-3}$ sec./eV for the maximal deficit in the neutrino flux, i.e., for $\beta \simeq 30$.

Improved experimental precision may reveal these effects of physics beyond standard model, including NSIs of neutrinos.
With the advancement in the Chameleon detection techniques, we would be hopeful these signals to be clearly appeared in near future terrestrial experiments (see also Ref. \cite{Burrage} for a more recent review and discussion of the various feasible experimental tests on such a screening mechanism).
Afterwards, we conclude that the effects of the decay process on neutrino flux will be confirmed as a sub-leading reason for the well-known solar neutrino problem.
Furthermore, data analysis of future neutrino flavor conversion measurements, as has already been done by neutrino experiments such as SK \cite{Super-Kamiokande}, HK \cite{Hyper-Kamiokande} and JUNO \cite{JUNO}, determining the different mixing parameters in the framework of NSI might improve the values of related parameters.

\appendix
\numberwithin{equation}{section}
\makeatletter
\newcommand{\section@cntformat}{Appendix \thesection:\ }
\makeatother

\section{Chameleon screening mechanism}\label{app1}
The Chameleon scalar field is coupled to the different matter species through an exponential conformal factor (\ref{eqn3}) with an inverse-power law potential
\begin{eqnarray}\label{eqnA.1}
\mathcal{V}(\phi)= M^{4+n}\phi^{-n}.
\end{eqnarray}
Here, $\beta$ is the scalar-matter coupling constant, $n$ is a positive number, and $M$ is the parameter of mass scale.

\begin{figure}[H]
	\centering
	\includegraphics[scale=1]{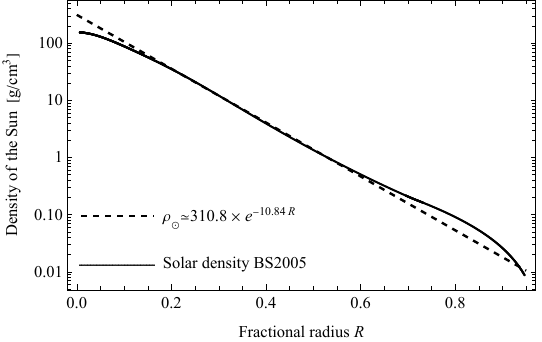}
	\caption{\footnotesize{The solar matter density profile is drawn in terms of the fractional radius for the BS05 model \cite{Bahcall-Density} (solid line) and its corresponding exponential fitting (dashed line).}}
	\label{fig8}
\end{figure}

Variation from the action (\ref{eqn2}) with respect to the field $\phi$ gives the equation of motion for non-relativistic matter
\begin{eqnarray}\label{eqnA.2}
\square \phi = \mathcal{V}_{\text{eff.},\phi},
\end{eqnarray}
where $\mathcal{V}_{\text{eff.},\phi}(\phi) = \mathcal{V}_{,\phi}(\phi) + A_{,\phi}(\phi) \rho$ is the effective potential, which governs the dynamics of the scalar field.
The most critical ingredient in the Chameleon field model is that the minimum of the effective potential, i.e.,
\begin{eqnarray}\label{eqnA.3}
\phi_{\text{min.}} = \bigg[\frac{n M^{4+n} M_p}{\beta \rho}\bigg]^{\frac{1}{n+1}},
\end{eqnarray}
and the mass of fluctuations around the minimum, i.e.,
\begin{eqnarray}\label{eqnA.4}
\begin{split}
&m^2_{\text{min.}} = \frac{n(n+1) M^{4+n}}{\phi^{n+2}_{\text{min}}} + \frac{\rho \beta^2}{M_p^2},
\end{split}
\end{eqnarray}
both depend explicitly on the ambient matter density $\rho$.
Expanding the scalar field around its background value up to linear order, i.e., $\phi(r) = \phi_0 + \delta\phi(r)$, allows us to reduce Eq.(\ref{eqnA.2}) to an equation for the perturbative part
\begin{eqnarray}\label{eqnA.5}
\frac{d^2 \delta\phi}{dr^2} + \frac{2}{r} \frac{d\delta\phi}{dr}  -  m_{\text{min}}^2 (\phi_0) \delta\phi = \frac{\beta(\phi_0)}{M_p} \rho(r),
\end{eqnarray}
where the source $\rho(r)$ on the right-hand side imposes matter effects in the scalar field. 
The solar standard model allows us to investigate the interior density profile \cite{Bahcall-Density}. 
To have analytical solutions, density profile inside the Sun can be approximated by a best-fit exponential function (see Fig. \ref{fig8}):
\begin{eqnarray}\label{eqnA.6}
\rho(R) =
\begin{cases}
\rho_c e^{-\lambda R_{\odot} R} & \left(R\leq 1\right)\\
\rho_0 & \left(R>1\right)
\end{cases}
,
\end{eqnarray}
where $R_{\odot}$ and $R\equiv r/R_{\odot}$ are the solar and dimensionless fractional radii, respectively.
This exponential fit can also be used in analyses of matter effects on solar neutrino propagation, where the constant parameters $\lambda = 10.84 R^{-1}_\odot \simeq 1.56 \times 10^{-5}$ km$^{-1}$ and $\rho_c \simeq 310.8~\text{g/cm}^3$ are obtained from best-fitting.

\begin{figure}[H]
	\centering
	\includegraphics[scale=0.60]{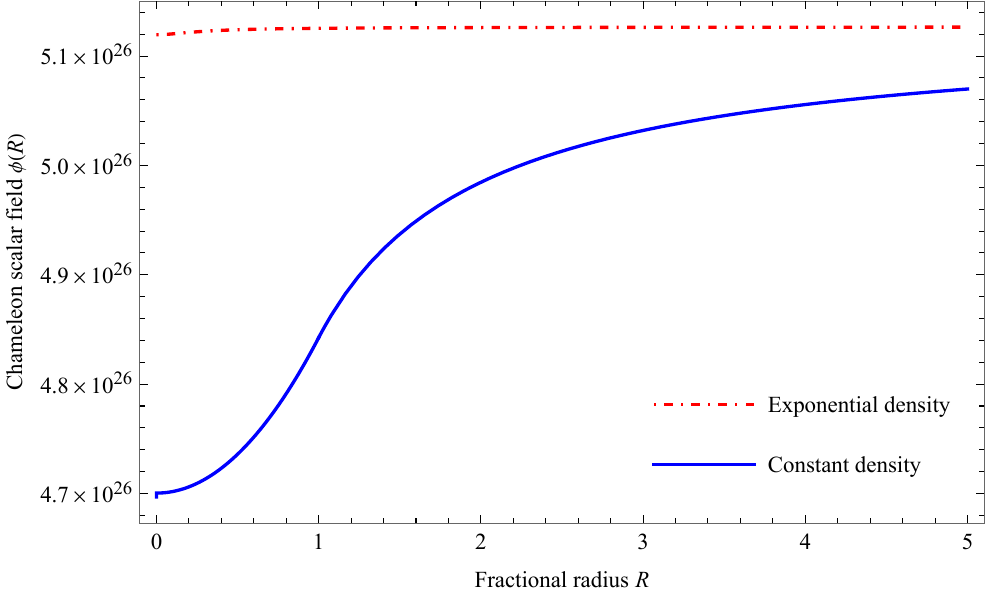}
	\caption{\footnotesize{This figure shows how the scalar field is affected by the matter density $\rho(R)$. We have assumed that $\beta \sim1$.}}
	\label{fig9}
\end{figure}

We can find a solution analytically to the scalar field equation (\ref{eqnA.5}) inside and outside the Sun placed in a medium with constant matter density $\rho_0$. Thus; we have
\begin{eqnarray}\label{eqnA.7}
\begin{split}
&\delta\phi_{\text{in}}(R) = \frac{e^{-(\lambda + 2m_{\text{in}})R_{\odot} R}}{2 m_{\text{in}} M_p R_{\odot} \left(m_{\text{in}}^2 - \lambda^2\right)^2 R} \bigg[C_1 M_p \left(m_{\text{in}}^2 - \lambda^2 \right)^2 \left(e^{2 m_{\text{in}} R_{\odot} R} - 1 \right) e^{(\lambda + m_{\text{in}}) R_{\odot} R} \\& ~~~~~~~~~~~+ 2 \beta  m_{\text{in}}  \rho_c  \left(e^{2 m_{\text{in}} R_{\odot} R} \left(2 \lambda + \left(\lambda^2 - m_{\text{in}}^2\right) R_{\odot} R\right) - 2 \lambda e^{(\lambda + m_{\text{in}}) R_{\odot}R}\right)\bigg], &&(R<1)
\end{split}
\end{eqnarray}
and
\begin{eqnarray}\label{eqnA.8}
\begin{split}
\delta\phi_{\text{out}}(R) = C_2 \frac{e^{- m_{\text{out}} R_{\odot} R}}{R}. && &&(R>1)
\end{split}
\end{eqnarray}
To obtain the integration constants, we apply the boundary conditions
\begin{eqnarray}\label{eqnA.9}
\begin{split}
& \frac{d\delta\phi}{dR} = 0  &&\text{at} && R\longrightarrow 0,
\\& \delta\phi \longrightarrow 0 &&\text{at} && R\longrightarrow \infty,
\end{split}
\end{eqnarray}
and the continuity conditions at the solar surface $R=1$.
By assuming $m_{\text{in}} R_{\odot} \gg 1$ and $m_{\text{out}} R_{\odot} \ll 1$, we fix these two constants as follows:
\begin{eqnarray}\label{eqnA.10}
\begin{split}
&C_1 = \frac{2 \beta  \rho_c e^{- 2 m_{\text{in}} R_{\odot}}}{M_p (m_{\text{in}}^2 -\lambda^2)^2} \bigg[e^{(m_{\text{in}} -\lambda)R_{\odot}} \left(m_{\text{in}}^2 (1 - \lambda  R_{\odot}) + \lambda \left(\lambda - m_{\text{out}} (\lambda R_{\odot} + 2) + \lambda^2 R_{\odot}\right)\right) - 2 \lambda m_{\text{in}}\bigg],
\end{split}
\end{eqnarray}
and
\begin{eqnarray}\label{eqnA.11}
\begin{split}
&C_2 = \frac{\beta \rho_c e^{-(2m_{\text{in}} + \lambda)R_{\odot}}} {M_p m_{\text{in}} R_{\odot} (m_{\text{in}}^2 - \lambda^2)^2}   \bigg[ -m_{\text{in}}^3 R_{\odot} e^{2 m_{\text{in}} R_{\odot}} - m_{\text{in}}^2 (\lambda R_{\odot} - 1) e^{2 m_{\text{in}} R_{\odot}} \\& ~~~+ \lambda^2 (\lambda R_{\odot} + 1) e^{2 m_{\text{in}} R_{\odot}} + \lambda m_{\text{in}} \left((\lambda R_{\odot} + 2) e^{2 m_{\text{in}} R_{\odot}} - 4 e^{(\lambda + m_{\text{in}}) R_{\odot}}\right)\bigg].
\end{split}
\end{eqnarray}
As a particular case of interest, by setting $\lambda \rightarrow 0$, all above relations can be re-written for constant or slowly varying matter density $\rho_c$.
Now, we can plot the treatment of the scalar field in terms of the dimensionless fractional radius $R$. Figure \ref{fig9} shows the Chameleon scalar field for $\beta\sim 1$ and two various cases: the constant and exponential densities. As can be seen, both curves show different behaviors inside the Sun but tend to the same asymptotic value outside it.
On the other side, we can see how the coupling parameter $\beta$ can impact the Chameleon scalar field. As shown in Fig.{\ref{fig10}, the field $\phi$ chooses smaller asymptotic values when $\beta$ increases.
\begin{figure}[H]
	\centering
	\includegraphics[scale=0.50]{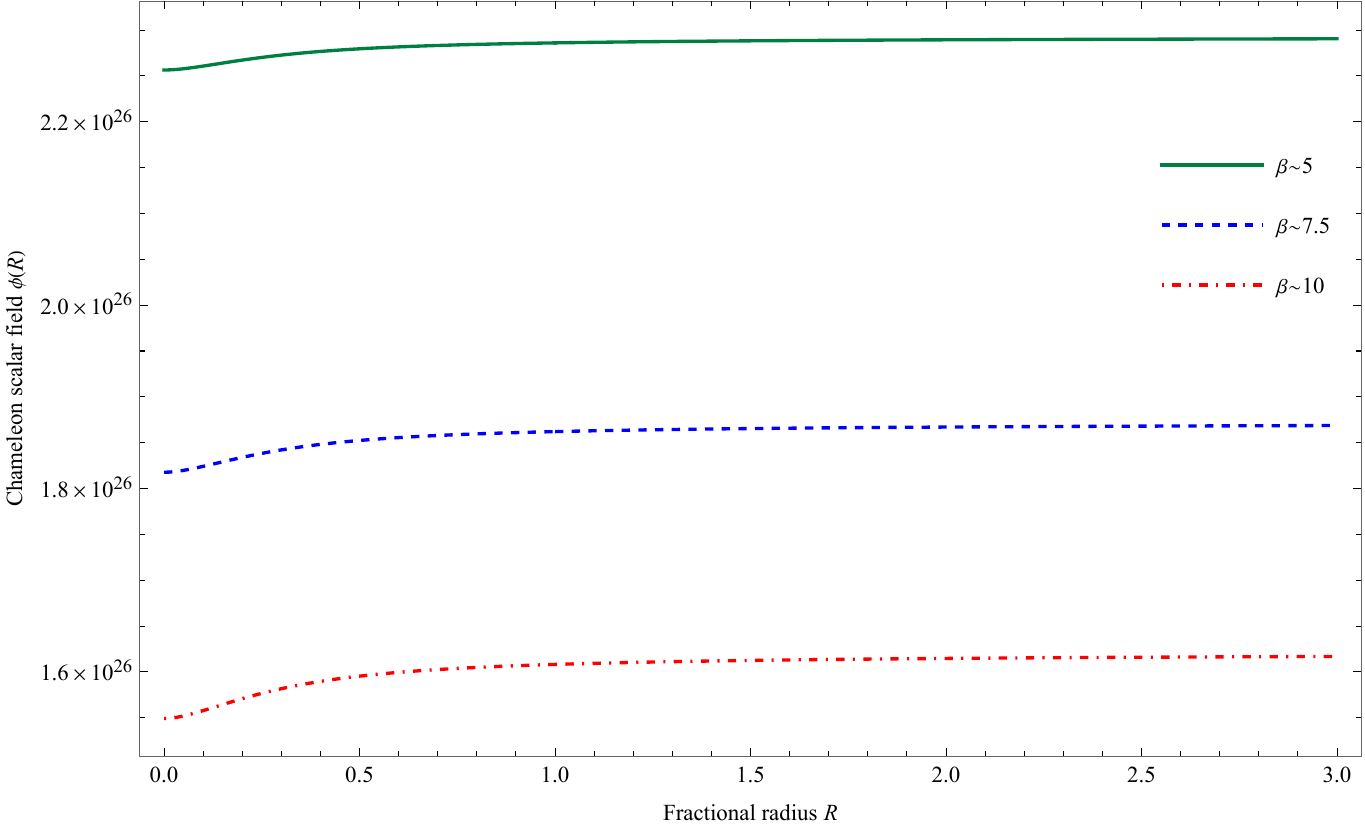}
	\caption{\footnotesize{The Chameleon scalar field for three different values of the coupling strengths $\beta \in \{5,7.5,10\}$. The field, in each case, approaches an asymptotic value outside the body, which increases with decreasing $\beta$. The field values are all in eV.}}
	\label{fig10}
\end{figure}

In Fig. \ref{fig11} we also show the numerical solution corresponding to the Chameleon model parameter $M$ as given in the legends with $\beta \sim1$ for the coupling parameter devoted to the gravitational strength. As can be seen in this case the numerical solution outside the Sun approaches to the asymptotic values.
\begin{figure}[H]
	\centering
	\includegraphics[scale=0.50]{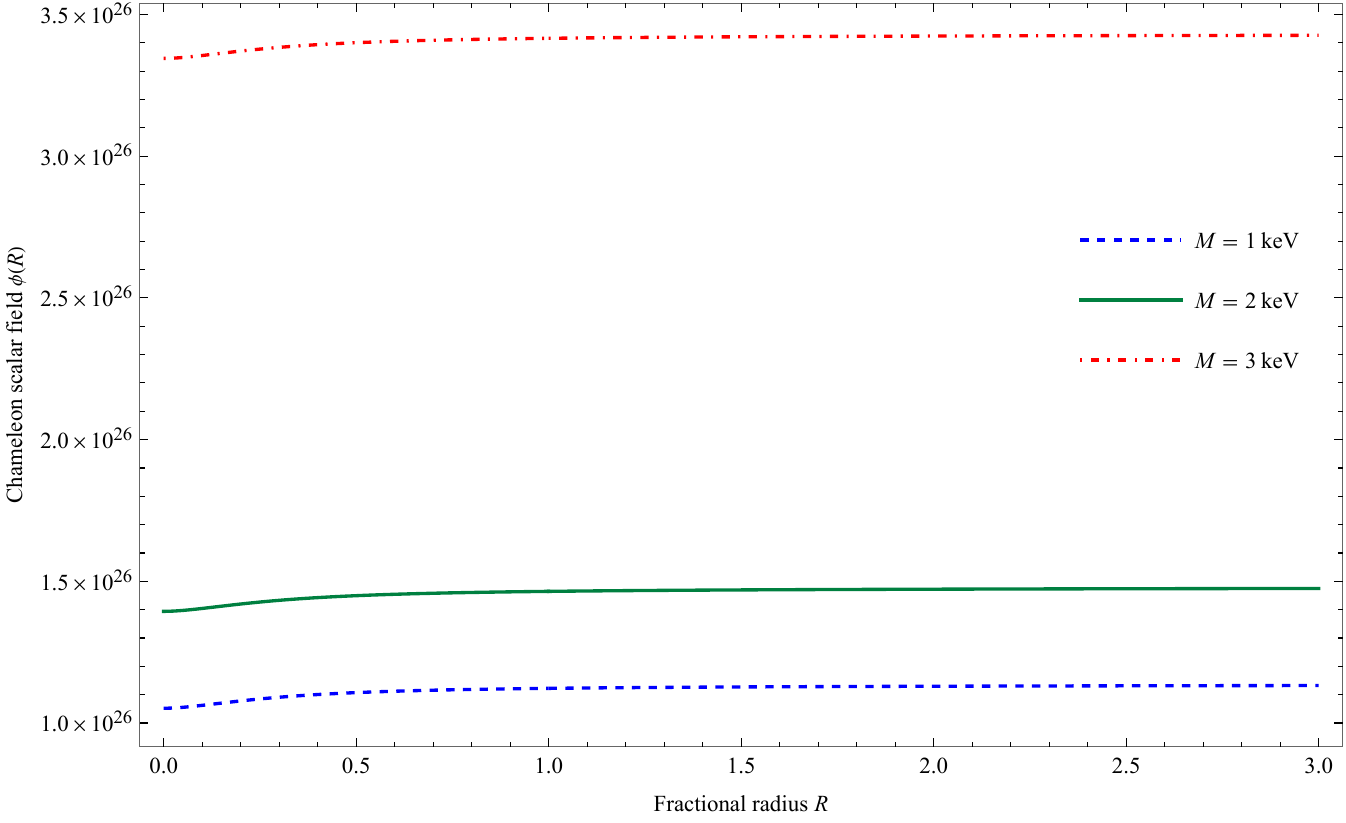}
	\caption{\footnotesize{The field profile in terms of the fractional radius $R$ and for various mass-scale parameters $M$. This case corresponds to $\beta \sim1$. All field profiles are in eV.}}
	\label{fig11}
\end{figure}

\end{document}